\documentclass[prb,twocolumn,showpacs,letterpaper,superscriptaddress]{revtex4}%
\usepackage{amsfonts}
\usepackage{amsmath}
\usepackage{amssymb}
\usepackage{graphicx}
\usepackage[colorlinks]{hyperref}
\providecommand{\U}[1]{\protect\rule{.1in}{.1in}}

\begin{document}
\title{Nonequilibrium spin transport on Au(111) surfaces}
\author{Ming-Hao Liu}\thanks{Present address: %
No. 2-1, Fushou Lane, Chengsiang Village, Gangshan Township,
Kaohsiung County 82064, Taiwan}%
\email{d92222010@ntu.edu.tw}%
\affiliation{Department of Physics, National Taiwan
University, Taipei 10617, Taiwan}
\author{Son-Hsien Chen}
\affiliation{Department of Physics, National Taiwan University,
Taipei 10617, Taiwan}%
\affiliation{Department of Physics and Astronomy, University of
Delaware, Newark, DE 19716-2570, USA}%
\author{Ching-Ray Chang}
\affiliation{Department of Physics, National Taiwan University, Taipei 10617, Taiwan}
\pacs{73.20.At, 73.23.-b, 71.70.Ej}
\begin{abstract}
The well-known experimentally observed \textit{sp}-derived Au(111)
Shockley surface states with Rashba spin splitting are perfectly fit
by an effective tight-binding model, considering a two-dimensional
hexagonal lattice with $p_{z}$-orbital and nearest neighbor hopping
only. The extracted realistic band parameters are then imported to
perform the Landauer-Keldysh formalism to calculate nonequilibrium
spin transport in a two-terminal setup sandwiching a Au(111) surface
channel. Obtained results show strong spin density on the Au(111)
surface and demonstrate (i) intrinsic spin-Hall effect, (ii)
current-induced spin polarization, and (iii) Rashba spin precession,
all of which have been experimentally observed in semiconductor
heterostructures, but not in metallic surface states. We therefore
urge experiments in the latter for these spin phenomena.
\end{abstract}
\date{\today}
\maketitle

\section{Introduction}

Two-dimensional electron gas (2DEG) is known to exist in various systems,
including semiconductor heterostructures \cite{Davies98} and metallic surface
states.\cite{Davison92} Due to the lack of inversion symmetry introduced by
the interface or surface, the spin degeneracy, the combining consequence of
the time reversal symmetry (Kramers degeneracy) and the inversion symmetry, is
removed and the energy dispersion becomes spin-split. In semiconductor
heterostructures, one of the underlying mechanisms leading to such spin
splitting is known as the Rashba spin-orbit coupling,\cite{Bychkov84} which
stimulates a series of discussion on plenty of intriguing spin-dependent
phenomena. Well studied phenomena include spin
precession,\cite{Kato04a,Crooker05} spin-Hall effect
(SHE),\cite{Kato04,CISPexp2,CISPexp4} and current-induced spin polarization
(CISP),\cite{CISPexp1,CISPexp2,CISPexp3,CISPexp4} all of which have been
experimentally observed in semiconductor heterostructures. Contrarily, none of
these in metallic surface states is reported, even though the Rashba effect
has been shown to exist therein.\cite{Lashell96,Bihlmayer06}

To the lowest order in the inplane wave vector $k_{\parallel},$ the two
spin-split energy branches are expressed as $E_{\pm}=\hbar^{2}k_{\parallel
}^{2}/2m^{\star}\pm\alpha k_{\parallel}$ ($m^{\star}$ the electron effective
mass), so that the Rashba spin splitting $\Delta E=E_{+}-E_{-}=2\alpha
k_{\parallel}$ is linear in $k_{\parallel}$. Here the proportional constant
$\alpha$ is commonly referred to as the Rashba coupling constant or the Rashba
parameter. Typical values of $\alpha$ in semiconductor heterostructures are at
most of the order of $10^{-2}$ $%
\operatorname{eV}%
\operatorname{\text{\AA}}%
$,\cite{CISPexp2,CISPexp3,Nitta97} while in metallic surface states $\alpha$
can be one or two orders larger.

The first evidence of spin splitting in metallic surface states was pioneered
by LaShell \textit{et al.} on Au(111) surfaces at room
temperature.\cite{Lashell96} The origin of their observed spin splitting was
later recognized as the Rashba effect by performing the first-principles
electronic-structure and photoemission calculations,\cite{Henk03,Henk04} which
are in good agreement with the spin-resolved photoemission
experiments.\cite{Henk04,Hoesch04} Concluded Rashba parameter of the Au(111)
surface states is about $\alpha=0.36$ $%
\operatorname{eV}%
\operatorname{\text{\AA}}%
$. Subsequent findings of giant Rashba spin-orbit coupling is also claimed in
Bi(111) surfaces\cite{Koroteev04}\ with $\alpha\approx0.83$ $%
\operatorname{eV}%
\operatorname{\text{\AA}}%
$ and in Bi/Ag(111) surface alloy\cite{Ast07} with $\alpha\approx3.05$ $%
\operatorname{eV}%
\operatorname{\text{\AA}}%
$.

It is therefore legitimate to expect the previously mentioned spin-dependent
phenomena to be observed on those metallic surfaces with strong Rashba
coupling. In this paper we theoretically study nonequilibrium spin transport
in 2DEG held by Au(111) surface states, which exhibit not only strong Rashba
coupling but also simple parabola-like dispersions.\cite{Reinert03} The latter
characteristic enables successful description of the band structure using the
simplest tight-binding model (TBM), which then provides the Landauer-Keldysh
formalism (LKF)\cite{Datta95,Nikolic05,Nikolic06} with reasonable or even
realistic band parameters.\begin{figure}[b]
\centering
\includegraphics[width=8.6cm]{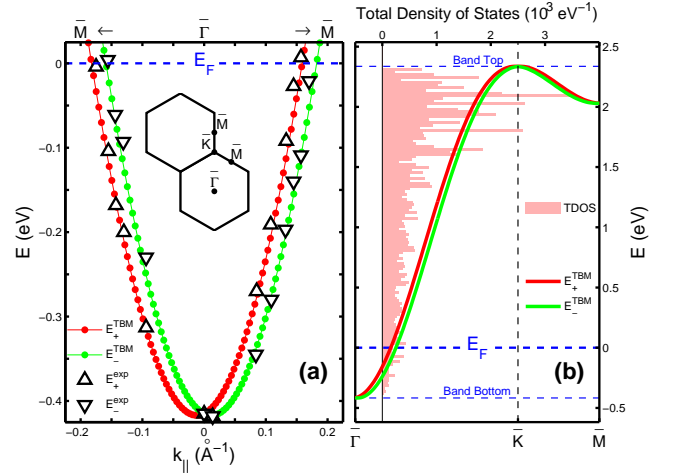} \caption{{}(Color online) (a)
Tight-binding energy dispersion $E_{\pm}^{\text{TBM}}$ and the experimentally
measured binding energy $E_{\pm}^{\text{exp}}$ of Ref. \onlinecite{Lashell96},
along the $\bar{\Gamma}\bar{M}$ direction. The surface Brillouin zone is
sketched in the inset. (b) Total density of states and $E_{\pm}^{\text{TBM}}$
along $\bar{\Gamma}\bar{K}$ and $\bar{K}\bar{M}$ directions.}%
\label{fig1}%
\end{figure}

This paper is organized as follows. In Sec. \ref{sec2} we describe the Au(111)
surface band structure by using an effective TBM, through which the
experimentally measured energy dispersions\cite{Lashell96} can be perfectly
reproduced. Section \ref{sec3} is devoted to nonequilibrium spin transport on
a finite Au(111) surface channel attached to two external leads, using the LKF
with band parameters extracted in Sec. \ref{sec2}. The intrinsic SHE, the
CISP, and the Rashba spin precession will be shown by directly imaging the
local spin densities. We conclude in Sec. \ref{sec4}.

\section{A\lowercase{u}(111) surface band structure\label{sec2}}

\subsection{Effective tight-binding model}

We first demonstrate that the \textit{sp}-derived Shockley surface states on
Au(111) from Ref. \onlinecite{Lashell96}\ can be well described by an
effective TBM [see Fig. \ref{fig1}(a)] for a single sheet of two-dimensional
hexagonal lattice, taking into account only $p_{z}$-orbital hopping between
nearest neighbors, subject to the Rashba spin-orbit coupling. The Hamiltonian
matrix can be written as\cite{Grosso00,Kane05}%
\begin{equation}
\mathbb{H}=E_{p}\openone+\sum_{\mathbf{t}_{I}}e^{i\mathbf{k}_{\parallel}%
\cdot\mathbf{t}_{I}}\left[  V_{pp\pi}\openone+V_{R}\mathbf{e}_{z}\cdot\left(
\vec{\sigma}\times\mathbf{t}_{I}\right)  \right]  , \label{H}%
\end{equation}
where $\openone$ is the $2\times2$ identity matrix, $E_{p}$ is the $p$-orbital
energy, $\mathbf{t}_{I}$ represents the six nearest neighbor hopping vectors,
$V_{pp\pi}$ is the band parameter describing the orbital integral under the
two-center approximation of Slater and Koster,\cite{Slater54} $V_{R}$ is the
Rashba hopping parameter, and $\vec{\sigma}=\left(  \sigma_{x},\sigma
_{y},\sigma_{z}\right)  $ is the Pauli matrix vector. The three terms in Eq.
(\ref{H}) are the energy band offset, the kinetic hopping, and the Rashba
hopping, respectively. Arranging the two primitive translation vectors for the
hexagonal lattice as $\mathbf{t}_{1}=(\sqrt{3}/2,1/2,0)a$ and $\mathbf{t}%
_{2}=(-\sqrt{3}/2,1/2,0)a$ where $a$ is the lattice constant, the six nearest
neighbor hopping vectors are $\mathbf{t}_{I}=\pm\mathbf{t}_{1},\pm
\mathbf{t}_{2},\pm(\mathbf{t}_{1}+\mathbf{t}_{2}),$ and Eq. (\ref{H}) then
takes the explicit form of%
\begin{equation}
\mathbb{H}(\mathbf{k}_{\parallel})=\left(
\begin{array}
[c]{cc}%
E_{p}+G(\mathbf{k}_{\parallel}) & F(\mathbf{k}_{\parallel})\\
F^{\ast}(\mathbf{k}_{\parallel}) & E_{p}+G(\mathbf{k}_{\parallel})
\end{array}
\right)  \label{Hmatrix}%
\end{equation}
with%
\begin{align}
F(\mathbf{k}_{\parallel})  &  =iV_{R}[(1+\sqrt{3}i)\sin\mathbf{k}_{\parallel
}\cdot\mathbf{t}_{1}\nonumber\\
&  +(1-\sqrt{3}i)\sin\mathbf{k}_{\parallel}\cdot\mathbf{t}_{2}+2\sin
k_{y}a]\label{F}\\
G(\mathbf{k}_{\parallel})  &  =2V_{pp\pi}[2\cos\frac{\sqrt{3}k_{x}a}{2}%
\cos\frac{k_{y}a}{2}+\cos\left(  k_{y}a\right)  ]. \label{G}%
\end{align}
Equation (\ref{Hmatrix}) can be diagonalized to yield the energy dispersions
\begin{equation}
E(\mathbf{k}_{\parallel})=E_{p}+G(\mathbf{k}_{\parallel})\pm|F(\mathbf{k}%
_{\parallel})|. \label{E}%
\end{equation}
Noting from Eq. (\ref{F}) that $V_{R}$ is embedded in $F(\mathbf{k}%
_{\parallel})$, the above dispersion contains the Rashba term to all (odd)
orders in $k_{\parallel}$.

In the vicinity of $\bar{\Gamma},$ i.e., $|\mathbf{k}_{\parallel}|a\ll1,$ Eqs.
(\ref{F}) and (\ref{G}) are approximated by $F(\mathbf{k}_{\parallel}%
)\approx-3V_{R}\left(  k_{x}-ik_{y}\right)  a$ and $G(\mathbf{k}_{\parallel
})\approx6V_{pp\pi}-(3V_{pp\pi}a^{2}/2)k_{\parallel}^{2}$, respectively, and
the Hamiltonian matrix (\ref{Hmatrix}) then takes the form%
\begin{equation}
\mathbb{H}_{k_{\parallel}a\ll1}=\left(
\begin{array}
[c]{cc}%
E_{0}-\dfrac{3V_{pp\pi}a^{2}}{2}k_{\parallel}^{2} & -3V_{R}\left(
k_{x}-ik_{y}\right)  a\\
-3V_{R}\left(  k_{x}+ik_{y}\right)  a & E_{0}-\dfrac{3V_{pp\pi}a^{2}}%
{2}k_{\parallel}^{2}%
\end{array}
\right)  , \label{Hmatrix2}%
\end{equation}
where $E_{0}\equiv E_{p}+6V_{pp\pi}.$ Equation (\ref{Hmatrix2}) is consistent
with the $p_{z}$-resolved effective Hamiltonian of the earlier TBM by Petersen
and Hedeg\aa rd, who considered all the three $p$-orbitals, subject to the
intra-atomic spin-orbit coupling.\cite{Petersen00}

\subsection{Extraction of band parameters}

We now fit our tight-binding dispersions (\ref{E}) with the experiment of Ref.
\onlinecite{Lashell96}. This can be done by comparing the low-$k_{\parallel}$
expansion of Eq. (\ref{E}),%
\begin{equation}
\left.  E(\mathbf{k}_{\parallel})\right\vert _{ka\ll1}\approx E_{p}+6V_{pp\pi
}-\frac{3V_{pp\pi}a^{2}}{2}k_{\parallel}^{2}\pm3V_{R}a|\mathbf{k}_{\parallel
}|, \label{Elowk}%
\end{equation}
with that of the free-electron model, $E(k_{\parallel})=E_{0}+(\hbar
^{2}/2m^{\star})k_{\parallel}^{2}\pm\alpha k_{\parallel}.$ In addition to the
band offset $E_{0}=E_{p}+6V_{pp\pi}$, we identity $V_{pp\pi}=-(2/3a^{2}%
)(\hbar^{2}/2m^{\star})$ and $V_{R}=\alpha/3a$. Using the reciprocal vector
$\mathbf{g}_{1}=(4\pi/\sqrt{3}a)(1/2,\sqrt{3}/2,0)$ and $\bar{M}$ $=1.26$ $%
\operatorname{\text{\AA}}%
^{-1}$ from Ref. \onlinecite{Lashell96}, we have $\hbar^{2}/2m^{\star}%
\approx15.2$ $%
\operatorname{eV}%
\operatorname{\text{\AA}}%
^{2},$ $\alpha\approx0.3557$ $%
\operatorname{eV}%
\operatorname{\text{\AA}}%
$, and $E_{0}\approx-0.415$ $%
\operatorname{eV}%
$. The norm of $\mathbf{g}_{1}$ gives $\bar{M}$ such that the lattice constant
is $a=4\pi/\sqrt{3}\left\vert \mathbf{g}_{1}\right\vert =5.\,\allowbreak7581$
$%
\operatorname{\text{\AA}}%
$. Hence we deduce $V_{pp\pi}=-0.3056$ $%
\operatorname{eV}%
$, $V_{R}=0.0206$ $%
\operatorname{eV}%
$, and $E_{p}=1.4188$ $%
\operatorname{eV}%
$. Substituting these parameters into Eqs. (\ref{F})--(\ref{E}), a nearly
perfect consistency between our effective TBM and the experimentally measured
binding energy of Ref. \onlinecite{Lashell96} can be seen in Fig.
\ref{fig1}(a). The experimentally measured Fermi surface of the concentric
rings slightly distorted from circles\cite{Reinert03} can be reproduced as
well, but we do not explicitly show.

\section{Nonequilibrium spin transport\label{sec3}}

\subsection{Landauer-Keldysh formalism vs tight-binding model}

Next we apply the Landauer-Keldysh formalism,\cite{Nikolic05} namely the
nonequilibrium Keldysh Green's function formalism\cite{Keldysh65} applied on
Landauer multiterminal ballistic nanostructures. For detailed introduction to
the LKF, see Refs. \onlinecite{Datta95,Nikolic06}. To make use of the
previously extracted band parameters in the LKF calculation, we consider the
second-quantized single particle Hamiltonian,\cite{Kane05}%
\begin{equation}
\mathcal{H}=\sum_{n}\varepsilon_{n}c_{n}^{\dag}c_{n}+\sum_{\left\langle
m,n\right\rangle }c_{m}^{\dag}\left[  t_{0}+it_{R}\left(  \vec{\sigma}%
\times\mathbf{d}_{mn}\right)  _{z}\right]  c_{n}, \label{Hsecond}%
\end{equation}
which is equivalent to Eq. (\ref{H}), provided $\varepsilon_{n}=E_{p}$,
$t_{0}=-V_{pp\pi}$, and $t_{R}=V_{R}$. In Eq. (\ref{Hsecond}), $c_{n}^{\dag}$
($c_{n}$) is the creation (annihilation) operator of the electron on site $n$,
$\langle m,n\rangle$ means that sites $m$ and $n$ are nearest neighbors to
each other, and $\mathbf{d}_{mn}$ is the unit vector pointing from $n$ to $m$.
Despite the different system sizes TBM and LKF consider (infinite for TBM and
finite for LKF) and different functions they provide (simple band calculation
by TBM and nonequilibrium transport by LKF), the equivalence of the underlying
Hamiltonians should contain the same physics. The explicit correspondence can
be shown by comparing the band structure by the TBM with the total density of
states (TDOS) by the LKF, provided that the same parameters are used.

For the LKF calculation, we consider a $80(a\sqrt{3}/2)\times11a\approx398.9$
$%
\operatorname{\text{\AA}}%
\times63.3$ $%
\operatorname{\text{\AA}}%
$ (total number of sites $N=931$) channel made of an ideal Au(111) surface, in
perfect contact with two unbiased normal metal leads at the left and right
ends of the sample. We will further image the nonequilibrium spin transport on
this two-terminal setup later. As shown in Fig. \ref{fig1}(b), the range of
the calculated nonvanishing TDOS is consistent with the TBM dispersion along
the $\bar{\Gamma}\bar{K}$ direction, which corresponds to the nearest neighbor
hopping direction as we considered in the underlying Hamiltonian
(\ref{Hsecond}).\begin{figure}[t]
\centering
\includegraphics[width=8.6cm]{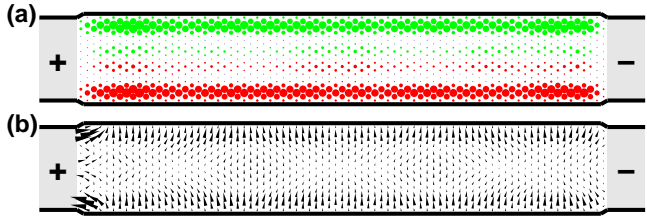} \caption{{}(Color online) Local spin
density of (a) the out-of-plane component $\langle S_{z}\rangle$ and (b) the
inplane component $(\langle S_{x}\rangle,\langle S_{y}\rangle)$ in a $398.9$
$\times63.3$ (total number of sites $N=931$) conducting sample made of Au(111)
surface. The size of each local marker depicts the magnitude. In (a), red/dark
(green/light) dots denote $\langle S_{z}\rangle>0$ ($\langle S_{z}\rangle<0$).
The maximum of $\langle S_{z}\rangle$ is $1.2\times10^{-3}\left(
\hbar/2\right)  $ while the mean of $\langle S_{y}\rangle$ is $3.16\times
10^{-4}(\hbar/2)$.}%
\label{fig2}%
\end{figure}

\subsection{Injection of unpolarized current: Intrinsic spin-Hall effect and
current-induced spin polarization}

Combination of the consistency between the experimental and the TBM
dispersions, and that between the dispersion by the TBM and the TDOS by the
LKF, indirectly demonstrates that the following imaging of local spin
densities by the LKF stands on an experimental footing. As a first
demonstration of the nonequilibrium spin transport, we turn on the bias of
potential difference $eV_{0}=0.2$ $%
\operatorname{eV}%
$ between the two normal metal leads. We will denote $\pm eV_{0}/2$ on the
leads by $\pm$ sign. With such injection of an unpolarized electron current,
we expect (i) the SHE of the intrinsic type, and (ii) the CISP, which follows
the Rashba eigenspin direction of the lower energy branch. Both of these can
be seen respectively in Figs. \ref{fig2}(a) and \ref{fig2}(b). The former
shows an antisymmetric out-of-plane spin accumulation with $\langle
S_{z}\rangle_{\max}=1.2\times10^{-3}(\hbar/2)$ at lateral edges, while the
latter shows that the inplane components of spins mostly point to $+y$
axis\cite{Liu08} with average value $\overline{\langle S_{y}\rangle
}=3.16\times10^{-4}(\hbar/2)$. Note that the local spin densities shown here
represent, by definition, the site-dependent total number of
spins.\cite{Nikolic06} Dividing $\overline{\langle S_{y}\rangle}$ by the
hexagonal unit cell area $\sqrt{3}a^{2}/2$, we deduce that the obtained spin
(area) density due to CISP is in average $1101$ $%
\operatorname{\mu m}%
^{-2}$, which is clearly much stronger than that observed in the CISP
experiment of Ref. \onlinecite{CISPexp1}, where the spin (volume) density less
than $10$ $%
\operatorname{\mu m}%
^{-3}$ (corresponding to a even weaker spin area density) is reported.

\subsection{Injection of spin-polarized current: Rashba spin precession}

Next we inject spin-polarized currents by replacing the left (source) lead
with a ferromagnetic electrode. Previously, the self-energy due to the normal
metal lead, which is assumed to be semi-infinite, in thermal equilibrium, and
in perfect contact with the sample, can be obtained by solving the surface
Green's function of the lead,\cite{Datta95} subject to Hamiltonian
$\mathcal{H}_{\text{lead}}=\mathbf{p}^{2}/2m+V$. The momentum operator
$\mathbf{p}=\left(  p_{x},p_{y}\right)  $ is two-dimensional and the potential
$V$ describes a infinite potential well of a semi-infinite rectangle shape.
The exact form of the lead self-energy reads%
\begin{align}
\Sigma^{R}\left(  p_{1},p_{2}\right)   &  =-\frac{2t}{N_{d}+1}\sum
_{n=1}^{\infty}\sin\frac{n\pi p_{1}}{N_{d}+1}\sin\frac{n\pi p_{2}}{N_{d}%
+1}\nonumber\\
&  \times\openone\frac{e^{ik_{n}a}\sin\left(  k_{n}a\right)  }{k_{n}a}
\label{SE}%
\end{align}
with $k_{n}a=\sqrt{\left(  E-E_{n}\right)  /t}$ and $E_{n}=[n\pi
/(N_{d}+1)]^{2}t.$ Here $p_{1(2)}=1,2,\cdots,N_{d}$ is the lateral position
(implicitly in units of lattice constant $a$) of the edge site in the sample
in contact with the lead, $t$ is the coupling between the sample and the lead
and is usually set equal to the kinetic hopping $t_{0}$ in the sample, and
$\left(  N_{d}+1\right)  a$ is implicitly assumed to be the width of the lead.

To take into account the exchange field inside the ferromagnetic lead, we
adopt the Weiss mean field approximation and add a Zeeman term $-\mu_{B}%
\vec{\sigma}\cdot\mathbf{B}_{ex}$ ($\mu_{B}\approx5.8\times10^{-5}%
\operatorname{eV}%
\operatorname{T}%
^{-1}$ the Bohr magneton) to $\mathcal{H}_{\text{lead}}$. Typical exchange
field may be as high as $|\mathbf{B}_{ex}|\approx10^{-3}$ $%
\operatorname{T}%
$ (Ref. \onlinecite{Kittel05}), leading to $\left\vert \mu_{B}\mathbf{B}%
_{ex}\right\vert \approx5.8\times10^{-2}$ $%
\operatorname{eV}%
$. We will take this value in the forthcoming spin precession demonstration.
The explicit form of the self-energy is obtained by substituting in Eq.
(\ref{SE}) $k_{n}a\rightarrow k_{n}^{\sigma}a=\sqrt{\left(  E-E_{n}%
+\sigma\left\vert \mu_{B}\mathbf{B}_{ex}\right\vert \right)  /t}$ and
$\openone\rightarrow\sum_{\sigma=\pm}|\mathbf{n}\sigma\rangle\langle
\mathbf{n}\sigma|$, where $|\mathbf{n}\sigma\rangle$ is the spin-1/2 state ket
with quantization axis $\mathbf{n}$\textbf{.}\cite{Sakurai94}\begin{figure}[t]
\centering
\includegraphics[width=8.6cm]{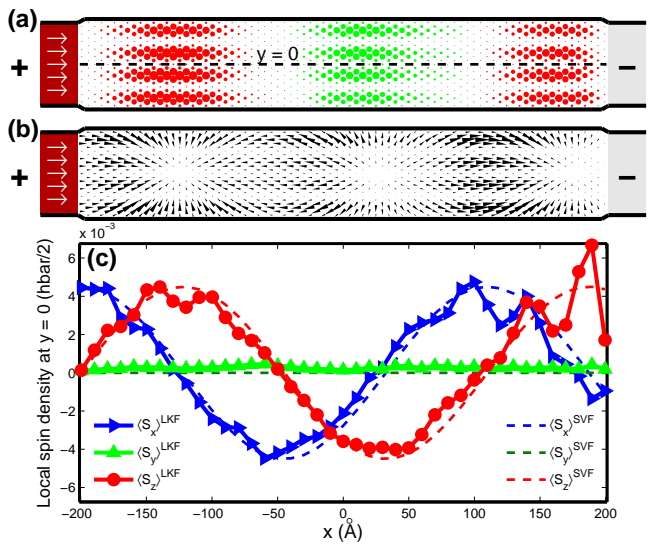} \caption{{}(Color online) Local spin
density of (a) the out-of-plane component $\langle S_{z}\rangle$ and (b) the
inplane component $(\langle S_{x}\rangle,\langle S_{y}\rangle),$ in a
conducting sample made of Au(111) surface, subject to a ferromagnetic source
lead with $+x$ magnetization. (c) $\langle S_{x}\rangle,$ $\langle
S_{y}\rangle,$ and $\langle S_{z}\rangle$ as a function of $x$ at $y=0$, i.e.,
along the dashed line sketched in (a). Computed values are compared with the
previously obtained spin vector formula based on quantum mechanics.}%
\label{fig3}%
\end{figure}

Applying the same voltage difference of $0.2$ $%
\operatorname{V}%
$ and magnetizing the ferromagnetic source lead along $+x$ axis, Figs.
\ref{fig3}(a) and \ref{fig3}(b) show the out-of-plane and inplane components
of the local spin densities, respectively. The injected $x$-polarized spins
moving along $+x$ and encountering the Rashba effective magnetic field
pointing to $-y$, are forced to precess about $-y$-axis counterclockwise, and
hence the Rashba spin precession is observed. In the free electron model, the
spin precession length (the distance within which the spin completes a
precession angle of $\pi$) is $L_{so}=(\pi/\alpha)(\hbar^{2}/2m^{\star
})\approx134$ $%
\operatorname{\text{\AA}}%
$. Thus the channel length $398.9$ $%
\operatorname{\text{\AA}}%
$ is about 3 times $L_{so}$, which is consistent to what we observe in Figs.
\ref{fig3}(a) and \ref{fig3}(b). Note that here the SHE competing with the
spin precession is relatively weak due to the strong exchange field we
consider in the source lead. However one can still observe the tiny asymmetry
of the $\langle S_{z}\rangle$ pattern of Fig. \ref{fig3}(a) along the lateral
direction (more $+\langle S_{z}\rangle$ and $-\langle S_{z}\rangle$
accumulations near the bottom and top edges, respectively). In the following
we will concentrate on the spin precession only.

To compare the LKF results with the free electron model in further detail, we
recall the spin vector formula [see Eq. (6) of Ref. \onlinecite{Liu06c}],
which takes the form of $(\langle S_{x}\rangle,\langle S_{y}\rangle,\langle
S_{z}\rangle)=(\cos\Delta\theta,0,\sin\Delta\theta)$ here with $\Delta
\theta=(2m^{\star}/\hbar^{2})\alpha x=2.34\times10^{-2}x$, where $x$ is in
unit of $%
\operatorname{\text{\AA}}%
$. Note that a factor of $\sqrt{3}/2$ responsible for the net and actual
hopping distances has to be taken into account in $x$, since the crystal
structure information remains. Accordingly, good agreement between the LKF and
the spin vector formula can be seen in Fig. \ref{fig3}(c). Note that in view
of both Figs. \ref{fig2} and \ref{fig3}, size and edge effects, arising from
the charge distribution, are also observed. The former, the size effect,
appears in the modulation along $y$ direction with roughly 4 peaks
corresponding to a wave length$\allowbreak\approx32%
\operatorname{\text{\AA}}%
$, roughly shorter than the Fermi wavelength $2\pi/k_{F}\approx38$ $%
\operatorname{\text{\AA}}%
$; the latter, the edge effect, appears in the abnormal charge accumulation
near the side and drain edges.

\section{Conclusion\label{sec4}}

In conclusion, we have shown that the \textit{sp}-derived Shockley surface
states on Au(111),\cite{Lashell96,Henk03,Henk04,Hoesch04,Reinert03} which
extend over the first few layers though, can be well described by an effective
TBM for a two-dimensional hexagonal lattice, taking into account $p_{z}%
$-orbital and nearest neighbor hopping only. Required parameters in the
nonequilibrium spin transport calculation by the LKF, demonstrating (i)
intrinsic SHE and CISP due to injection of unpolarized current and (ii) the
Rashba spin precession due to injection of spin-polarized current, thus stand
on an experimental footing of the pioneering work of LaShell \textit{et.
al.}.\cite{Lashell96} Calculated local spin densities in all the three spin
phenomena are much stronger than those in semiconductor heterostructures.
Whereas the magnetic optical Kerr effect (MOKE) can sensitively detect a spin
volume density of less than 10 spins per $%
\operatorname{\mu m}%
^{3}$ (Ref. \onlinecite{CISPexp1}), our results of more than $10^{3}$ spins
per $%
\operatorname{\mu m}%
^{2}$ suggest definitely measurable nonequilibrium spin transport supported by
the Au(111) surface states and others with even stronger Rashba coupling such
as Bi(111) surfaces\cite{Koroteev04}\ or Bi/Ag(111) surface
alloys.\cite{Ast07} Last, in addition to the stronger local spin densities
induced by stronger Rashba coupling, the spin precession length (typically of
the order of 1 $%
\operatorname{\mu m}%
$ in semiconductor heterostructures) in these surface states is greatly
reduced [$134$ $%
\operatorname{\text{\AA}}%
$ for Au(111) reported here], such that the fine structure of the spin
patterns due to spin precession or the intrinsic SHE requires high resolution
apparatus such as the spin-polarized scanning tunneling
microscopy.\cite{Bode03}

\begin{acknowledgments}
Financial support of the Republic of China National Science Council Grant No.
95-2112-M-002-044-MY3 is gratefully acknowledged.
\end{acknowledgments}

\bibliographystyle{apsrev}
\bibliography{mhl2}

\end{document}